\documentclass[twocolumn,nofootinbib,amsmath,amssymb,a4paper]{revtex4}

\usepackage{graphicx}
\usepackage{dcolumn}
\usepackage{bm}

\newcommand{\lsim}{
\mathrel{\hbox{\rlap{\hbox{\lower4pt\hbox{$\sim$}}}\hbox{$<$}}}}

\newcommand{\gsim}{
\mathrel{\hbox{\rlap{\hbox{\lower4pt\hbox{$\sim$}}}\hbox{$>$}}}}
\def\MSbar{\hbox{\tiny ${\overline{\rm MS}}$}}

\begin{document}

\title{Inclusive B decays from resummed perturbation theory}

\author{Einan Gardi}

\affiliation{
Cavendish Laboratory, University of Cambridge, \\J J Thomson
Avenue, Cambridge, CB3 0HE, UK\\
{and}\\
Department of Applied Mathematics \& Theoretical Physics,\\
Wilberforce Road, Cambridge CB3 0WA,~UK
}

\begin{abstract}
 I review the recent progress in computing inclusive B decay widths and spectra and its implications on the interpretation of measurements from the B factories.
I discuss the inclusive charmless semileptonic decay, ${\bar B}\to X_u l\bar{\nu}$, which provides the most robust determination of the CKM parameter $|V_{\rm ub}|$, and the rare radiative decay, ${\bar B}\to X_s \gamma$, which constrains flavor violation beyond the Standard Model. I demonstrate that precise predictions for the experimentally--relevant branching fractions can be derived from resummed perturbation theory
and explain the way in which the resummation further provides guidance in parametrizing non-perturbative Fermi--motion effects.
Finally I address the comparison between theory and data and discuss future prospects.
\end{abstract}

\maketitle

\section{Introduction}

Inclusive B decay measurements have a central r\^ole in flavour physics. Inclusive semileptonic decays~\cite{HFAG_semileptonic} ${\bar B}\to X_{c,u} l \bar{\nu}$ provide the most accurate determinations of the CKM parameters $|V_{\rm cb}|$ and $|V_{\rm ub}|$, and thus strongly constrain the Unitarity Triangle~\cite{Charles:2004jd,Bona:2005eu}. This constraint is particularly important: being based on tree--level Weak decays it is totally insensitive to physics beyond the Standard Model (SM). This should be contrasted for example with the measurement of the ${\cal CP}$ violating phase $\sin (2\beta)$, which is sensitive to loop effects through $B^0- \bar{B}^0$ mixing, and can therefore be influenced by physics beyond the SM.
The potential to discover new physics crucially depends on our ability to make precise comparison between the two~\cite{Charles:2004jd,Bona:2005eu,Kagan:2000wm,Ball:2006xx,Ball:2006jz}.

Another major avenue in flavor physics is the use of rare decays to constrain flavor--changing neutral currents beyond the SM. Here the most stringent constraint is based on the inclusive Branching Fraction (BF) of radiative B decays, ${\bar B}\to X_s \gamma$,~see~\cite{HFAG_rad}.

\begin{figure}[tb]
\begin{center}
\includegraphics[width=0.54\textwidth]{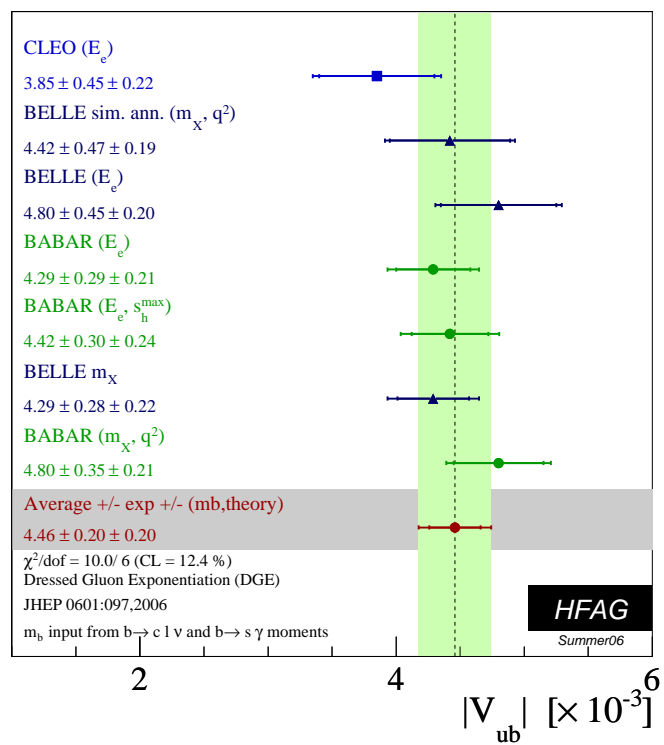}
\caption{HFAG world average $|V_{\rm ub}|$ using DGE: summary of all available inclusive measurements of ${\bar B}\to X_u l \bar{\nu}$ with different kinematic cuts on $M_X, \,q^2, E_l$ etc. In each case the partial BF was translated to a value of $|V_{\rm ub}|$ using DGE.   \label{fig:HFAG}}
\end{center}
\end{figure}

Our ability to exploit the potential the B factories depends on the accuracy at which we can compute the corresponding BF's. Since only part of the phase space is experimentally accessible,  theoretical calculations are needed not only for the total widths, but also for the spectra.

In ${\bar B}\to X_s \gamma$ and in ${\bar B}\to X_{u} l \bar{\nu}$ most events are characterized by jet--like kinematics, where the invariant mass of the hadronic system $m_X$ is small, much smaller than the energy ${\cal O}(m_b)$ that is released in the decay. Moreover, the measurements are restricted to this kinematic
region. Therefore, theoretical understanding of the limit $m_X\ll m_b$ becomes the key to the interpretation of the measurements.

A major part of this talk is devoted to the recent progress in our ability to compute inclusive decay spectra using resummed QCD perturbation theory~\cite{Gardi:2004ia,Gardi:2005yi,Andersen:2005bj,Andersen:2005mj,Andersen:2006hr}.
The most precise predictions for the experimentally--relevant BF's in ${\bar B}\to X_s \gamma$~\cite{Andersen:2005bj,Andersen:2006hr} and in ${\bar B}\to X_{u} l \bar{\nu}$~\cite{Andersen:2005mj} (see Fig.~\ref{fig:HFAG}) are now based on \emph{calculations} using resummed perturbation theory --- not anymore on a parametrization of experimental data using a ``shape function''. This breakthrough was achieved using Dressed Gluon Exponentiation (DGE)~\cite{DGE_review}, a resummation method that fully uses the inherent infrared--safety of the on-shell decay spectrum. I~will explain the conceptual differences between DGE and the ``shape function'' approach, show that definite predictions can be derived from perturbation theory despite its divergent nature, and finally explain the way in which the resummation provides guidance in parametrizing non-perturbative Fermi--motion effects. I will also discuss the determination of the ${\bar B}\to X_s \gamma$ BF and conclude by addressing the present theoretical uncertainty, the comparison between theory and data and future prospects.

\section{Breakthrough in computing decay spectra~\label{sec:DGE}}

The key to computing inclusive decay spectra is the understanding of the Sudakov region, where the hadronic system in the final state is \emph{jet--like}, namely it is characterized by a large hierarchy between the two lightcone momentum components, $P^{+}\ll P^{-}$, where $P^{\pm}\equiv E_X{\mp}|p_X|$ and $P^{-}$ is typically of the order of the heavy--quark mass.

\begin{figure}[tb]
\begin{center}
\includegraphics[width=0.39\textwidth,angle=90]{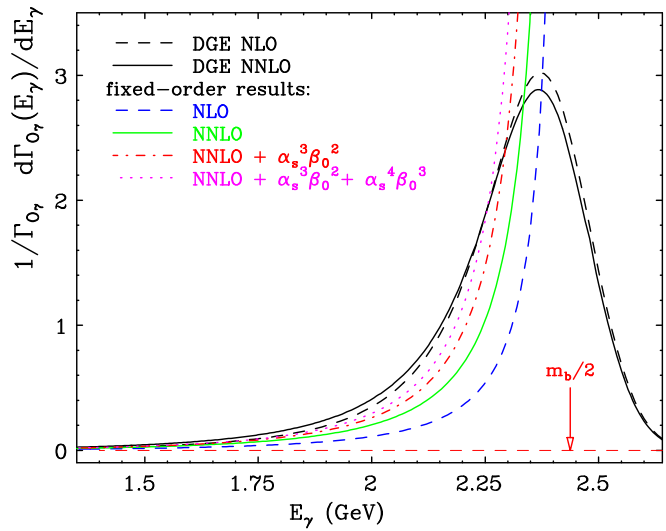}
\caption{The photon--energy spectrum in ${\bar B}\to X_s \gamma$ (for the electromagnetic dipole operators $O_7$) as computed by resummed perturbation theory (DGE) compared with fixed--order perturbation theory at NLO, NNLO~\cite{Melnikov:2005bx,Asatrian:2006sm} and higher order corrections in the large--$\beta_0$ limit~\cite{Gardi:2004ia}. The endpoint of the on-shell decay spectrum is $m_b^{\rm PV}/2$ (arrow) where the quark pole mass is $m_b^{\rm PV}=4.87$ GeV.  The endpoint of the physical spectrum is $M_B/2$ (end of the $E_{\gamma}$ axis) where the meson mass is $M_B=5.28$ GeV.  \label{fig:spectrum}}
\end{center}
\end{figure}

It has been understood long ago that the leading contribution to the spectrum for small $P^{+}$ involves the momentum distribution of the b quark in the meson~\cite{Bigi:1993ex,Neubert:1993um}. Indeed, if we imagine a hypothetical situation where the $b$ quark (of mass $m_b$) is on-shell inside the meson --- so it does not interact with the light--degrees--of--freedom carrying the residual energy $\bar{\Lambda}\equiv M_B-m_b$ ---
we would find that the spectrum, at any order, has support only for $p^+\equiv P^+-\bar{\Lambda}\geq 0$. In contrast, the physical spectrum has support for $P^+\geq 0$. Fig.~\ref{fig:spectrum} demonstrates these different support properties in the case of ${\bar B}\to X_s \gamma$ where $P^-=M_B$ and $P^+=M_B-2E_{\gamma}$. It is therefore clear that the spectrum\footnote{The physical spectrum near the exclusive limit depends also on the hadronic structure of the final--state jet (e.g. the ${\bar B}\to X_s \gamma$ spectrum contains the ${\bar B}\to \bar{K}^* \gamma$ peak) but inclusive observables such as the partial branching fractions and the moments (defined over a sufficiently large phase space) are insensitive to these details.} at small $P^+$ is dictated by the Fermi motion of the b quark inside the meson.

The conventional approach to describing the spectrum has been based on the analogy with deep inelastic structure functions: the non-perturbative momentum distribution function is parametrized and is convoluted with a hard coefficient function that is computed with an infrared cutoff.
In this approach the ansatz for the momentum distribution function essentially dictates the shape of the spectrum in the entire peak region, hence the name ``shape function''. This is unfortunate, as there is no theoretical guidance whatsoever as to how to parametrize this function. All that can be currently deduced from fits to data are the center of the ``shape function'' (the b quark mass in some definition) and its width. Even with much more precise data in the peak region, the tail of the ``shape function'' would not be well constrained. Thus, the prospects of improving the precision in this approach (beyond what has already been done~\cite{Bosch:2004th,Lange:2005yw}) are very limited.

A more promising avenue is that of Dressed Gluon Exponentiation (DGE)~\cite{DGE_review} where the first approximation to the spectrum is based on the resummed calculation of the on-shell decay spectrum; non-perturbative effects enter only as power corrections~\cite{Gardi:2004ia}.
DGE is a general resummation formalism~\cite{DGE_review,Gardi:2001di,Cacciari:2002xb,Gardi:2001ny} for inclusive distributions near a kinematic threshold\footnote{Here the threshold corresponds to the phase--space limit where the small lightcone momentum component vanishes, $p^+\to~0$.}. It goes beyond the standard Sudakov resummation framework by incorporating renormalon resummation in the calculation of the exponent. This has proven effective~\cite{Cacciari:2002xb,Gardi:2001ny} in extending the range of applicability of perturbation theory nearer to threshold and in identifying the relevant non-perturbative corrections.

The impact of resummation on the calculation of the ${\bar B}\to X_s \gamma$ spectrum is demonstrated in Fig.~\ref{fig:spectrum}. In fixed--order perturbation theory one finds very large corrections. The
parametrically--leading corrections near the partonic endpoint $E_{\gamma}\to m_b/2$ ($p^+\to 0$) are Sudakov logarithms. However, conventional Sudakov resummation with a fixed logarithmic accuracy yields a divergent expansion (see Fig.~1 in \cite{Gardi:2005mf}) as it misses out important running--coupling corrections. In contrast the DGE resummed spectrum yields a stable result, see Fig.~\ref{fig:spectrum}.

A remarkable difference between the fixed--order result and the resummed one is in the support properties. While the perturbative endpoint, at any order, is $E_{\gamma}=m_b/2$ (i.e. $p^+=0$), the resummed result extends into the non-perturbative regime and tends to zero near the physical endpoint $E_{\gamma}= M_B/2$ (i.e. $P^+= 0$). Thus, resummation makes a qualitative difference. In the following we briefly describe what is done to arrive at this result, and then proceed to explain how Fermi--motion effects are taken into account as power corrections.

When applying perturbation theory to inclusive decay spectra it proves useful to consider the moments with respect to the ratio between the lightcone momentum components, e.g. in the semileptonic decay we define:
\begin{widetext}
\begin{equation}
\label{mom}
\frac{d\Gamma_N^{b\to X_u l\bar{\nu}}(p^-,\,E_l)}{dp^-\,dE_l}\,\equiv \,\int_0^{p^-} {dp^+}\,\left(1-\frac{p^+}{p^-}\right)^{N-1} \frac{d\Gamma^{b\to X_u l\bar{\nu}}(p^+,\,p^-,\,E_l)}{dp^+\,dp^-\,dE_l},
\end{equation}
\end{widetext}
where the partonic lightcone momentum components $p^{\pm}$ are related to the hadronic ones by: $p^{\pm}=P^{\pm}-\bar{\Lambda}$, where $\bar{\Lambda}=M_B-m_b$ is the energy of the light--degrees--of--freedom in the meson.
Note that we consider the moments of the \emph{fully differential width}~\cite{Andersen:2005mj}: e.g. in the semileptonic case the moments remain differential with respect to the large lightcone component $p^-$ as well as the lepton energy $E_l$. This is essential for performing soft gluon resummation.

Owing to their inclusive nature the moments (\ref{mom}) are \emph{infrared safe}: for any moment $N$ there is an exact cancellation of infrared  singularities at $p^+\to 0$ between real and virtual diagrams.
Thus, instead of considering directly the singular limit $p^+\to 0$, one considers the large--$N$ limit. In this limit one identifies three  characteristic scales, and the moments \emph{factorize}~\cite{Korchemsky:1994jb,Bauer:2001yt,Bosch:2004th,Gardi:2004ia} to all orders as follows:
\begin{eqnarray}
\label{fact}
&&\frac{d\Gamma_N^{b\to X_u l\bar{\nu}}(p^-,\,E_l)}{dp^-\,dE_l}\,=
\\ &&\hspace*{12pt}H(p^-,\,E_l)\, \nonumber \underbrace{J\big({p^-}/{\sqrt{N}},\,\mu\big)\, S_{b}\big({p^-}/{N},\,\mu\big)}_{{\rm Sud}(p^-,\, N)}\, + \,{\cal O}(1/N),
\end{eqnarray}
where the factorization--scale ($\mu$) dependence cancels exactly in the product in the Sudakov factor ${\rm Sud}(p^-,\, N)$.
Factorization facilitates the resummation of Sudakov logarithms~\cite{Korchemsky:1994jb,Akhoury:1995fp,Bauer:2001yt,Bosch:2004th,Gardi:2004ia,Aglietti:2001br,Becher:2005pd}~(see also~\cite{Contopanagos:1996nh,Korchemsky:1988hd,Sterman:1986aj,Sen:1981sd}), the corrections that dominate the dynamics at large $N$. These originate in two distinct regions of phase space: the final--state jet $J\big({p^-}/{\sqrt{N}},\,\mu\big)$ and the
initial--state quark--distribution function $S_b\big(p^-/N,\mu\big)$. The latter subprocess depends on the softest scale, $p^-/N$, and is therefore the first place where non-perturbative corrections should be included. Indeed, the \emph{meson} decay is described by
\begin{eqnarray}
\label{fact_meson}
&&\frac{d\Gamma_N^{{\bar{B}\to X_u l\bar{\nu}}}(p^-,\,E_l)}{dp^-\,dE_l}\,=\\ \nonumber 
&&\hspace*{12pt}H(p^-,\,E_l)\, J\big({p^-}/{\sqrt{N}},\,\mu\big)\, S_{\bar{B}}\big({p^-}/{N},\,\mu\big) \,+\,{\cal O}(1/N),
\end{eqnarray}
where $S_{\bar{B}}\big(p^-/N,\mu\big)$ stands for
the quark distribution in the meson.
Owing to the Fermi motion, this function differs from its perturbative counterpart, the quark distribution in an on-shell quark $S_b\big(p^-/N,\mu\big)$, by power corrections $(\Lambda N/p^-)^k$, which can be resummed into a new non-perturbative function,
\begin{equation}
\label{meson_quark}
S_{\bar B}\big(p^-/N,\mu\big)\,=\,S_b\big(p^-/N,\mu\big)\,\times\,{\cal F}(p^-/N).
\end{equation}
Note that ${\cal F}(p^-/N)$, in contrast to $S_{\bar B}\big(p^-/N,\mu\big)$,  is independent of the factorization scale $\mu$.
In Eq.~(\ref{calF_amb}) below we shall parametrize ${\cal F}(p^-/N)$ based on the infrared sensitivity exposed by the resummation.
Finally, the spectrum is given by an inverse Mellin transform to (\ref{mom}):
\begin{widetext}
\begin{equation}
\label{inv_Mellin}
\frac{d\Gamma^{\bar{B}\to X_u l\bar{\nu}}(P^+,\,P^-,\,E_l)}{dP^+\,dP^-\,dE_l}
\,=\left. \int_{c-i\infty}^{c+i\infty} \frac{dN}{2\pi \, i}\,\left(1-\frac{p^+}{p^-}\right)^{-N}
\frac{1}{p^-}\,\frac{d\Gamma_N^{\bar{B}\to X_u l\bar{\nu}}(p^-,\,E_l)}{dp^-\,dE_l}
\right\vert_{p^{\pm}=P^{\pm}-\bar{\Lambda}}
\end{equation}
\end{widetext}
where the integration contour runs parallel to the imaginary axis, to the right of the singularities of the integrand.

We are now in a position to understand some of the fundamental differences between the approach described above and the one based on a ``shape function'':
\begin{itemize}
\item{} When considering the on-shell decay in moment space for $N\gg 1$ we begin by assuming that $p^-$ (i.e. $m_b$) is sufficiently large so that even the soft scale characterizing the quark distribution function, $p^-/N$, is in the perturbative regime:
    \hbox{$p^-/N\gg \Lambda$}.
In this limit the on-shell decay spectrum is computable and provides a systematic approximation to the meson decay spectrum.
This should be contrasted with the ``shape function'' approach where one
works in momentum space and considers directly the situation where $p^+={\cal O}(\Lambda)$, so the ``shape function'' is treated as non-perturbative from the beginning.
\item{} Consequently, we can compute more and parametrize less.
We \emph{compute} the quark distribution in an on-shell quark, $S_b\big(p^-/N,\mu\big)$, which accounts for the radiation off the heavy quark. This radiation puts the heavy quark slightly off its mass shell. Thus, the quark decays off its mass shell, despite the fact that it was initially assumed on-shell. Of course, physically there is no unique distinction between this acquired virtuality and the ``primordial'' virtuality, which is ``a property of the bound state''.
This corresponds to the renormalon ambiguity in the calculation of the quark distribution function, which serves as a probe of the non-perturbative dynamics.
Thus, eventually we need to parametrize the power corrections ${\cal F}(p^-/N)$ that make for the difference between the quark distribution in the meson $S_{\bar{B}}\big(p^-/N,\mu\big)$ and that in an on-shell quark. In contrast,  the ``shape function'' approach is based on parametrizing the \emph{entire} function $S_{\bar{B}}\big(p^-/N,\mu\big)$.
\end{itemize}

The functions $J\big({p^-}/{\sqrt{N}},\,\mu\big)$\, and $S_{b}\big({p^-}/{N},\,\mu\big)$ in (\ref{fact}) satisfy Sudakov evolution equations~\cite{Contopanagos:1996nh,Korchemsky:1988hd,Sterman:1986aj,Sen:1981sd,Korchemsky:1992xv,Gardi:2002xm,Gardi:2005yi} whose general, all--order solution can be formulated as a scheme--invariant Borel sum~\cite{Cacciari:2002xb,Gardi:2002xm,Gardi:2004ia,Andersen:2005mj,Andersen:2005bj}:
\begin{widetext}
\begin{align}
\label{Sud_DGE} {\rm Sud}(p^-,\,N)&=\exp\bigg\{\frac{C_F}{\beta_0}
\int_0^{\infty}\frac{du}{u}\,\left(\frac{\Lambda}{p^-}\right)^{2u}\,
\bigg[B_{\cal
S}(u)\Gamma(-2u)\,\left(\frac{\Gamma(N)}{\Gamma(N-2u)}-\frac{1}{\Gamma(1-2u)}\right)
\nonumber \\  &
\hspace*{190pt}-B_{\cal J}(u)\Gamma(-u)
\left(\frac{\Gamma(N)}{\Gamma(N-u)}-\frac{1}{\Gamma(1-u)}\right)\bigg]\bigg\}\,
,
\end{align}
\end{widetext}
where $B_{\cal S}(u)$ and $B_{\cal J}(u)$ are the Borel representations of the Sudakov anomalous dimensions of the quark distribution and the jet function, respectively.
The Sudakov exponent has renormalon singularities at integer and half integer values of~$u$, except where $B_{\cal S,\, J}(u)$  vanish. The corresponding ambiguities, whose magnitude is determined by the residues of the poles in~(\ref{Sud_DGE}), are enhanced at large $N$ by powers of $N$. They indicate the presence of non-perturbative power corrections with a similar $N$ dependence. These power corrections \emph{exponentiate} together with the logarithms.

\begin{figure}[htb]
\begin{center}
\vspace*{-13pt}
\includegraphics[width=0.475\textwidth]{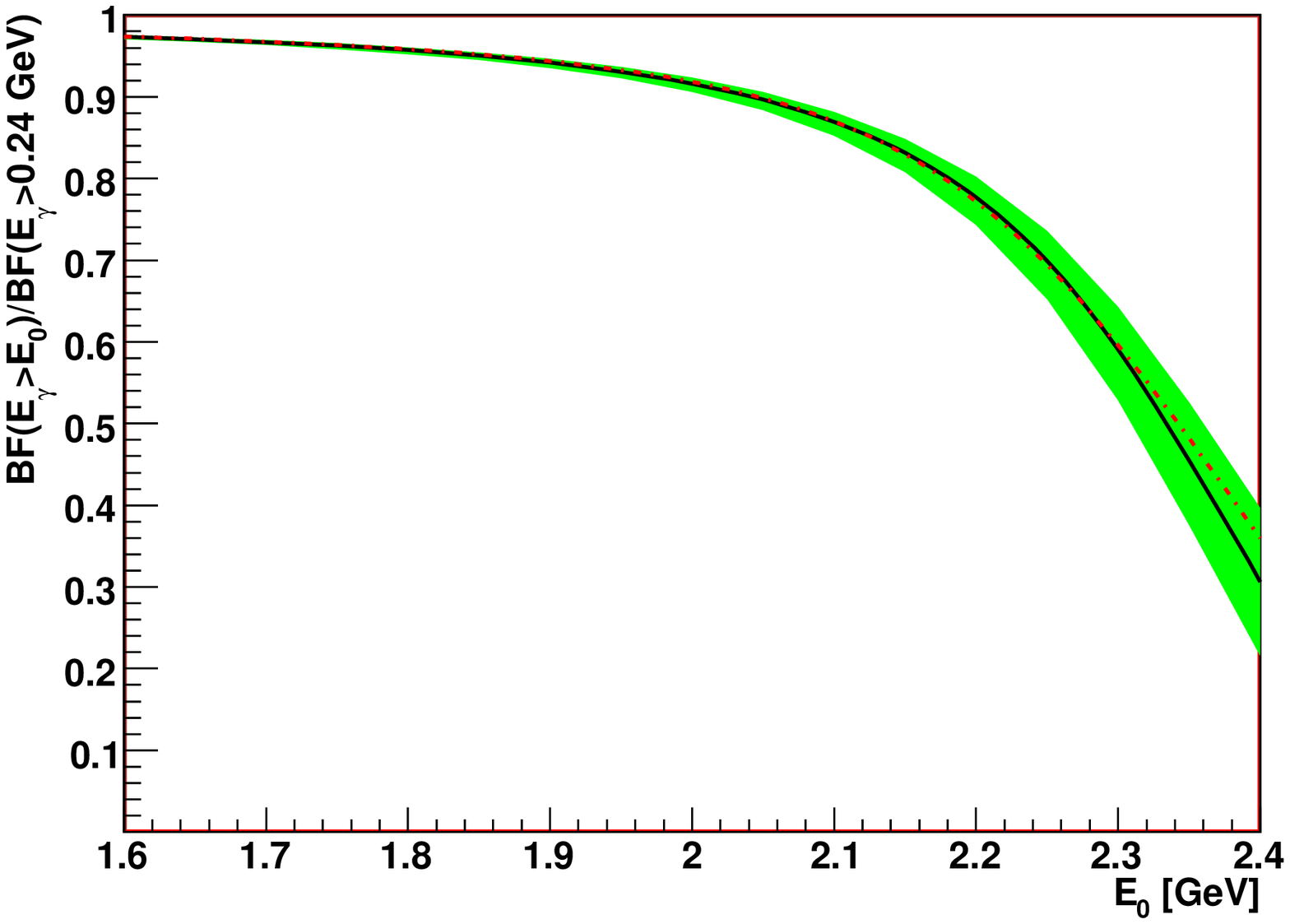}\\
\vspace*{-3pt}
\includegraphics[width=0.475\textwidth]{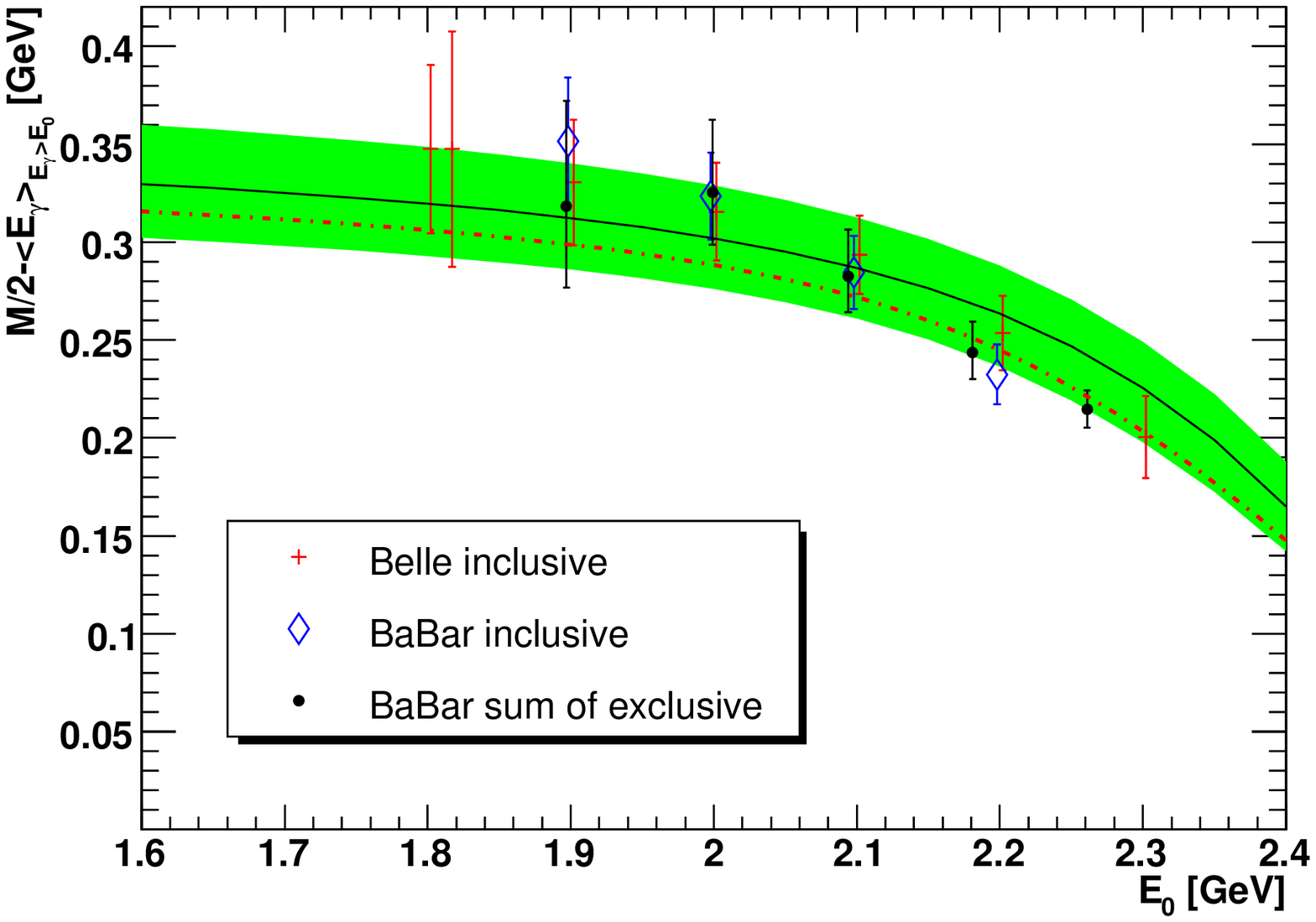}\\
\vspace*{-3pt}
\includegraphics[width=0.475\textwidth]{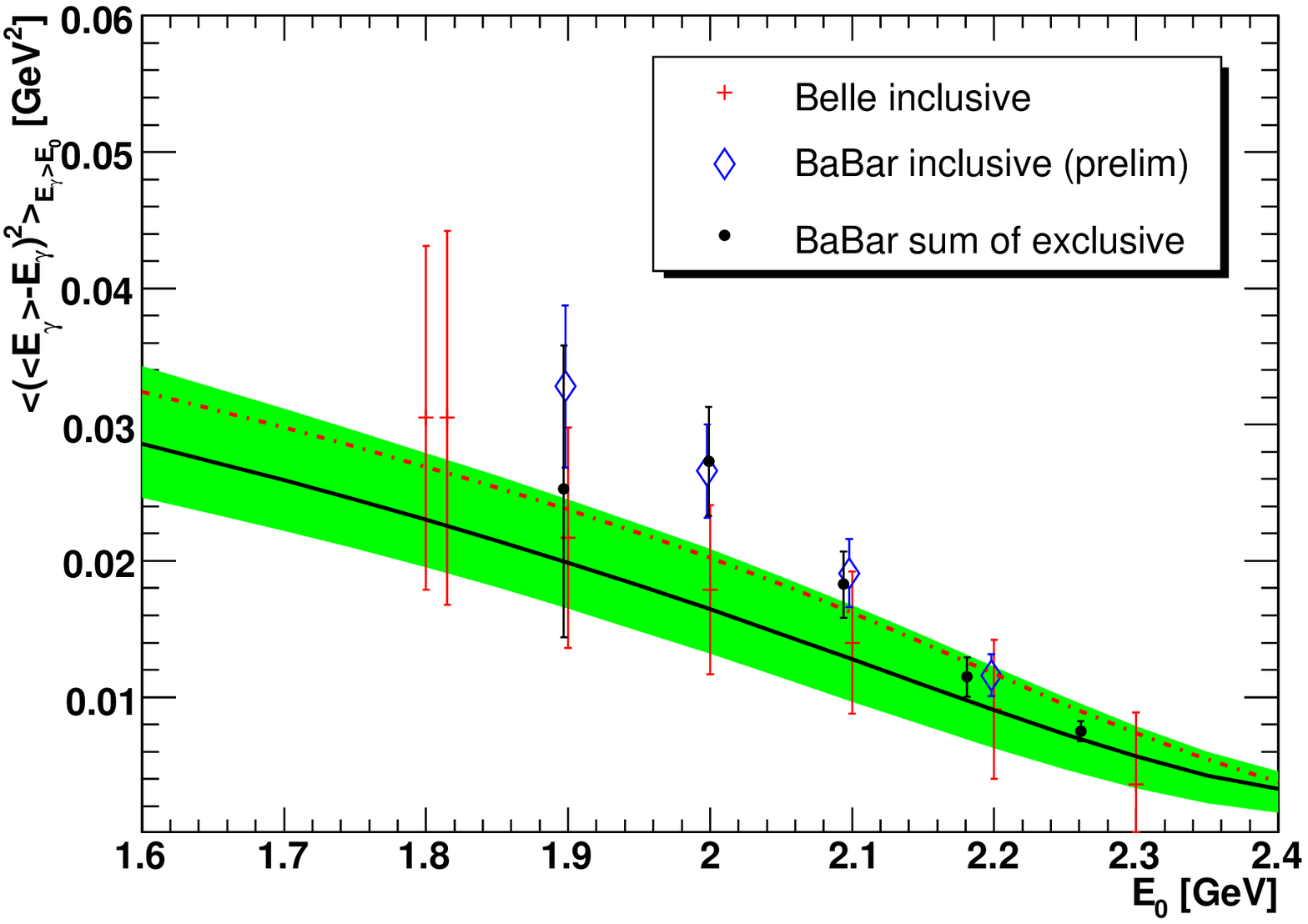}
\caption{Theoretical prediction for the partial BF as a function of the cut on the photon energy, $E_\gamma>E_0$, and the first two moments defined over the same range ((5.3) and (5.4) in \cite{Andersen:2005bj}). The moments are compared with available data.
No fits are done here. The effect of potential power corrections is illustrated by the difference between two curves: full line with $C_{3/2}=1; f_{\rm PV}=0$ and dotdashed one: $C_{3/2}=6.2; f_{\rm PV}=0.3$.
\vspace*{-25pt}
\label{fig:cut_moments}}
\end{center}
\end{figure}

The anomalous dimension functions $B_{\cal
S}(u)$ and $B_{\cal J}(u)$ are both known \cite{Gardi:2005yi} to NNLO, ${\cal O}(u^2)$. This facilitates Sudakov resummation with next--to--next--to--leading logarithmic accuracy~\cite{Andersen:2005bj} (see also \cite{Aglietti:2001br,Becher:2005pd}). In the DGE approach one takes a crucial step further~\cite{Gardi:2001ny,Gardi:2001di}: instead of expanding (\ref{Sud_DGE}) and truncating the expansion at a given logarithmic accuracy, the Borel integral in~(\ref{Sud_DGE}) --- the ``renormalon sum'' --- is performed in the Principal Value (PV) prescription, incorporating information on the behavior of the integrand away from the origin~\cite{Andersen:2005bj}. This information includes, in particular, the pattern of zeros of $B_{\cal S,\,J}(u)$, which one identifies using the large--$\beta_0$  limit: a zero implies the absence of a corresponding renormalon ambiguity in the exponent, a property we call ``infrared safety at the power level''~\cite{DGE_review}.

The renormalon ambiguities probe the infrared sensitivity and can therefore be used to parametrize non-perturbative power corrections.
Let us consider the power corrections associated with the quark distribution function. Upon choosing the PV prescription in the exponent in (\ref{Sud_DGE}) we essentially \emph{define} the quark distribution in an on-shell quark at the power level. In Eq. (\ref{meson_quark}), this amounts to a specific separation between perturbative and non-perturbative contributions to the (unique!) quark distribution in the meson. Operationally, promoting the resummed partonic calculation (\ref{fact}) into a prediction for the meson decay, (\ref{fact_meson}) to (\ref{inv_Mellin}), amounts to the following replacement:
\begin{align}
\begin{split}
{\rm Sud}(p^{-},N) &
\,\longrightarrow \,\left. {\rm
Sud}(p^{-},N)\right\vert_{\rm PV}\times {\cal F}_{\rm PV}(p^-/N)
\end{split}
\end{align}
where ${\cal F}_{\rm PV}(p^-/N)$ is constructed as a sum over the residues of the Borel integral (\ref{Sud_DGE})~\cite{Andersen:2006hr}:
\begin{eqnarray}
\label{calF_amb}
&&{\cal F}_{\rm PV}(p^-/N) =
\exp\bigg\{\frac{C_F}{\beta_0}\,\pi\,{f_{\rm PV}} \,
\,
 \times \\&&\hspace*{40pt}\nonumber
 \sum_{k=3}^{\infty} \,
\frac{(-1)^k}{k\, {k!}}\, \,{B_{\cal S}(k/2)}\left(\frac{\Lambda}{p^{-}}\right)^k {\Pi_{j=1}^k{(N-j)}}\bigg\}\,,
\end{eqnarray}
where we introduced a single dimensionless coefficient $f_{\rm PV}$ controlling the magnitude of the power corrections. On general grounds $f_{\rm PV}$ is expected to be of order 1: non-perturbative corrections are typically of the order of the ambiguity. Note that the sum in (\ref{calF_amb}) excludes the two leading powers $k=1$ and $k=2$. The leading renormalon ($k=1$) corresponds to the ambiguity in defining the pole mass and it cancels exactly~\cite{Gardi:2004ia} in the physical
spectrum~(\ref{inv_Mellin}). The second potential renormalon ($k=2$) does not appear~\cite{Gardi:2004ia}: this is an example of infrared safety at the power level. Consequently, the leading power correction corresponds to $k=3$ in (\ref{calF_amb}): it scales as $(N\Lambda/p^-)^3$. This high power makes the on-shell decay spectrum a good approximation to the physical meson decay.

It should be emphasized that although formally all power corrections in (\ref{calF_amb}) 
contribute at sufficiently large~$N$, because of the factorial suppression  only the first few powers are important. This facilitates the practical use of~(\ref{calF_amb}) despite the fact the behavior of $B_{\cal S}(k/2)$ for $k\geq 3$ is not known\footnote{$B_{\cal S}(u)$ is known analytically only in the large--$\beta_0$ limit.}. We describe
the Fermi--motion effects using two parameters~\cite{Andersen:2006hr} in (\ref{calF_amb}): the $k=3$ renormalon residue $B_{\cal S}(u=3/2)\equiv -0.23366\, C_{3/2}$ and the overall power correction coefficient $f_{\rm PV}$.
As shown in Fig.~\ref{fig:spectrum}, in the absence of power corrections, the support properties of the DGE spectrum are close, but not identical, to the physical ones. One can therefore constrain the parameters $\left(C_{3/2},\, f_{\rm PV}\right)$~\cite{Andersen:2006hr} using the support properties. Reassuringly, one finds that the typical magnitude of $f_{\rm PV}$ is $\lsim 1$, as expected on general grounds.

Fig.~\ref{fig:cut_moments} presents theoretical predictions for the
${\bar B}\to X_s \gamma$ BF as well as the average and the variance of the photon energy as a function of a cut $E_{\gamma}>E_0$.
The figure illustrates the potential effect of power corrections.
The total theoretical uncertainty is shown by the green band. A detailed analysis of the individual sources of uncertainty reveals that for any cut $E_0 < 2.2$ GeV, the uncertainty in the average energy is dominated by the parametric error assumed for the b quark mass ($m_b^{\MSbar}=4.20 \pm 0.04$); for $E_0 \geq 2.2$ GeV it is dominated by power corrections. The uncertainty in the variance is dominated by power corrections over the whole range $E_0\geq 1.8$ GeV. The conclusion is clear: a fit of the computed spectrum\footnote{A ${\tt \texttt{c}^{++}}$ code is available at: \\{\tt http://www.hep.phy.cam.ac.uk/$\sim$andersen/BDK/}} to the measured moments can yield an accurate determination of the b quark mass and the power corrections.
Fig.~\ref{fig:cut_moments} also presents the available data for the average energy and the variance. The data agrees well with the calculations over the entire range of cuts. This comparison suggests that the data is already precise enough to constrain the three parameters $(m_b,\, C_{3/2},\, f_{\rm PV})$.

Let us return now to the determination of $|V_{\rm ub}|$ from inclusive ${\bar B}\to X_u l\bar{\nu}$ measurements. The HFAG compilation in Fig.~\ref{fig:HFAG} summarizes the extracted values of $|V_{\rm ub}|$ from all available measurements based on different kinematics cuts. 
The crucial ingredient in this determination is the calculation of $R_{\rm cut}$, the partial BF within the region of measurement over the total $b\to u$ BF. The consistency  of $|V_{\rm ub}|$ between the different measurements
provides a good test of the theory predictions as well as the experimental analysis.
Nevertheless, some more focused analysis can directly test the error estimates and shed light on specific issues such as the relevance of Weak annihilation. Since the DGE calculation\footnote{
A ${\tt \texttt{c}^{++}}$ code is available at: \\
{\tt http://www.hep.phy.cam.ac.uk/$\sim$andersen/BDK/B2U/}}
is available at the level of the triple differential rate, any conceivable cut can be readily implemented. The calculation can thus be used to check the stability of the extracted $|V_{\rm ub}|$ as a function of the cut and, eventually to optimize the cuts.

We note that the present theoretical error in determining $|V_{\rm ub}|$ is comparable to the experimental one. Therefore, making good use of future data from the B factories requires more precise theoretical predictions.  Let us consider for example the case of the $M_X<1.7$ GeV cut, where a detailed
theoretical--error analysis has been done in~\cite{Andersen:2005mj}. The three dominant sources of uncertainty on $R_{\rm cut}(M_X<1.7)$  are: parametric error ($\sim 7\%$), higher--order corrections ($\sim 6\%$), and power corrections ($\lsim 3\%$). All three can be reduced in the foreseeable future: the parametric uncertainty is dominated by the short--distance b quark mass, the higher--order corrections are dominated by running--coupling effects which have been recently computed~\cite{Gambino:2006wk}, and the leading power corrections can be deduced from fits to the measured ${\bar B}\to X_s \gamma$ moments. Thus, the prospects for improving the determination of $|V_{\rm ub}|$ are high.

\section{The ${\bar B}\to X_s \gamma$ Branching Fraction}

The ${\bar B}\to X_s \gamma$ BF provides a crucial constraint on flavor--changing neutral currents beyond the SM. Experimental measurements by the B factories are getting more precise~\cite{HFAG_rad}:
\begin{eqnarray}
\label{radiative_exp}
{\cal B}\left({\bar B}\to X_s \gamma, \,E_{\gamma}>1.6\, {\rm GeV}\right)
= (355\pm 26)\cdot 10^{-6}\,.
\end{eqnarray}
Consequently, a very significant effort has been invested in the recent years in computing the SM BF with NNLO accuracy~(see e.g. \cite{Misiak:2006bw,Misiak:2006zs} and refs. therein). Despite the fact that only partial results are available --- and, in fact, very important ingredients such as the NNLO matrix element $G_{27}$ are missing ---
 first numerical estimates for the total BF have recently been published~\cite{Misiak:2006zs,Andersen:2006hr}:
\begin{eqnarray}
\label{radiative_result}
{\cal B}\left({\bar B}\to X_s \gamma, E_{\gamma}>1.6 {\rm GeV}\right)
\!=\! \left\{\!\!
\begin{array}{ll}
  (347\pm 48)\!\cdot\! 10^{-6} &\text{\cite{Andersen:2006hr}}\\
  (315\pm 23)\!\cdot\! 10^{-6} &\text{\cite{Misiak:2006zs}}
\end{array}\right. 
\end{eqnarray}
where the result of \cite{Andersen:2006hr} includes only $\beta_0\alpha_s^2$ terms~\cite{Bieri:2003ue} at NNLO while that of \cite{Misiak:2006zs} includes also partial results and estimates of other NNLO terms.
Although the two results in (\ref{radiative_result}) are consistent, the theoretical uncertainty is substantially different.
To have a more complete picture it is useful to compare the above estimates to NLO ones. Ours is
\begin{eqnarray}
\label{radiative_result_NLO}
\hspace*{-2pt}{\cal B}\left({\bar B}\to X_s \gamma, E_{\gamma}>1.6 {\rm GeV}\right)
\!=\! 
\begin{array}{ll}
  (313\pm 45)\!\cdot\! 10^{-6}\hspace*{-5pt} &\text{\cite{Andersen:2006hr}}\\
\end{array}
\end{eqnarray}
while the well known result by Gambino and Misiak from 2001 reads:
\begin{eqnarray}
\label{radiative_result_NLO_GM}
\hspace*{-2pt}{\cal B}\left({\bar B}\to X_s \gamma, E_{\gamma}>1.6 {\rm GeV}\right)
\!=\! 
\begin{array}{ll}
  (360\pm 30)\!\cdot\! 10^{-6}\hspace*{-5pt} &\text{\cite{Gambino:2001ew}}.
\end{array}
\end{eqnarray}
The various analysis differ in many ways, and a detailed comparison goes beyond the scope of this talk.
Here I would like to focus on one crucial aspect concerning the evaluation of the total BF, namely the way in which the b quark mass is treated. I will  then briefly discuss the cut dependence.

The dependence of the BF on the b quark mass is important since the mass enters the expression for the width at the fifth power:
\begin{widetext}
\begin{align}
\label{width}
\begin{split}
 &\Gamma(\bar{B}\to
X_s\gamma, E_{\gamma}>E_0) = \frac{\alpha_{\rm em} G_F^2}{32\pi^4}
\left|V_{\rm tb}V_{\rm ts}^*\right|^2  \,\left(m_b^{{\rm \overline{MS}}}(m_b)\right)^2
\,m_b^3
\sum_{i,j, \,i\leq
j}C_i(\mu)C_j(\mu)\, {G_{ij}(E_{0},\mu)}\\
 &\hspace*{20pt}= \frac{\alpha_{\rm em} G_F^2}{32\pi^4}
\left|V_{\rm tb}V_{\rm ts}^*\right|^2  \,\left(m_b^{{\rm \overline{MS}}}(m_b)\right)^2
\,{  m_b^3}\,
{
\underbrace{\bigg[f_0(\mu)+f_1(\mu)\frac{\alpha_s(\mu)}{\pi}+
f_2(\mu)\left(\frac{\alpha_s(\mu)}{\pi}\right)^2+\cdots\bigg]}_{{F}}},
\end{split}
\end{align}
\end{widetext}
where $C_i(\mu)$ are coefficient functions associated with different operators in the effective Weak Hamiltonian and $G_{ij}(E_{0},\mu)$ are the corresponding ${\bar B}\to X_s\gamma$ matrix elements. Here we made the important distinction between the short distance mass $m_b^{{\rm \overline{MS}}}$ and the pole mass $m_b$, which enter the expression for the width via the operator and via the phase--space integration, respectively.
Since the pole mass suffers from an ${\cal O}(\Lambda)$ renormalon ambiguity it cannot be directly used when evaluating the expression for the width at fixed order.
A similar problem is encountered in other decays, for example in the case of the charmless  semileptonic width one has:
\begin{widetext}
\begin{eqnarray}
\label{semileptonic_width}
 \Gamma\left({\bar B}\longrightarrow X_u
l\bar{\nu}, E_{\gamma}>E_0\right)=
\frac{G_F^2|V_{\rm ub}|^2{  m_b^5}}{192\pi^3}{  \underbrace{\bigg[1+s_1\frac{\alpha_s(\mu)}{\pi}+
s_2(\mu)\left(\frac{\alpha_s(\mu)}{\pi}\right)^2+\cdots\bigg]}_{{  G_u}}}.
\end{eqnarray}
\end{widetext}
Since this renormalon ambiguity cancels exactly~\cite{Bigi:1994em,Beneke:1994sw,Beneke:1994bc} if both the pole mass and the on-shell decay width ($G_u$ in (\ref{semileptonic_width}) and $F$ in (\ref{width})) are systematically considered to all orders, it does not necessarily limit the precision in evaluating the BF's. Two main strategies have been proposed: (1) the use of an alternative mass scheme, which is renormalon free; (2) Borel summation of the expansions for the pole mass and the decay width~\cite{Andersen:2005mj,Lee:2002px}. The first strategy is simple, as it can be applied at fixed order, however, it involves an uncontrolled uncertainty owing to the choice of the mass scheme.
This strategy was used in \cite{Gambino:2001ew,Misiak:2006zs} to obtain the results quoted above.
The second procedure requires renormalon resummation, but it has two important advantages: it deals directly with the most important corrections, those associated with the running coupling, and it facilitates using additional information on the large--order asymptotic behavior of the expansion. This approach has been used in the recent calculation of the total charmless semileptonic width~\cite{Andersen:2005mj}.
In the case of the radiative decay, however, there is an additional complication owing to the renormalization of the operators and their mixing. This gives an advantage to the fixed--order treatment where the scale dependence can be used as an uncertainty measure. Instead of using
some arbitrary mass scheme one can normalize the radiative decay width using the semileptonic width as follows~\cite{Andersen:2006hr}:
\begin{widetext}
\begin{align}
\begin{split}  \Gamma(\bar{B}\longrightarrow
X_s\gamma)& = \frac{\alpha_{\rm em} G_F^2}{32\pi^4}
\left|V_{\rm tb}V_{\rm ts}^*\right|^2
\,\left(m_b^{{\rm \overline{MS}}}(m_b)\right)^2
\times\hspace*{-40pt}{  \underbrace{{  m_b^3}\,\, {  G_u^{3/5}}}_{\rm Resummed:\, prescription\,\, independent}} \hspace*{-40pt}\times\bigg[{  F}/{  G_u^{3/5}}\bigg]_{\rm Fixed\,\,
Order},
\end{split}
\end{align}
\end{widetext}
where each of the factors is both renormalization--group invariant and renormalon--free: the product $m_b^3 G_u^{3/5}$ can be taken directly from the semileptonic decay analysis, while the function $F/G^{3/5}$ is simply treated at fixed order. This new strategy was used in \cite{Andersen:2006hr} to obtain the results quoted in (\ref{radiative_result_NLO}) and (\ref{radiative_result}) above. It turns out that in this formulation, and with only $\beta_0\alpha_s^2$ terms at NNLO, the renormalization scale dependence is still over $10\%$.

Our main conclusion is that the theoretical uncertainty on the total BF is not yet significantly smaller than at NLO. One reason for that is the large cancellation between different contributions to the width, e.g. between the $G_{77}$ sector, which is known in full to NNLO~\cite{Blokland:2005uk,Asatrian:2006ph,Melnikov:2005bx,Asatrian:2006sm}, and the $G_{27}$ one which is not.
It is quite clear that the uncertainty would indeed reduce upon completion of the NNLO calculation. Finally, comparing different strategies to deal with the b quark mass is absolutely essential for having a reliable error estimate.

Let us turn now to the dependence of the partial BF on the photon--energy cut. As discussed in Sec.~\ref{sec:DGE} the DGE approach opens the way for making quantitative predictions for the spectrum. As shown in Fig.~\ref{fig:cut_moments} there is good agreement between the calculation and the measurements of the first and second moments of the photon energy over the \emph{entire} range of cuts, $1.8 \leq E_0\leq 2.3$. Moreover, assuming similar cuts, the relative importance of non-perturbtaive corrections in the partial branching fraction is lower than in the moments. 
This gives us confidence in the predictions for the partial BF.

\section{Conclusions}

Over the past few years we made significant progress in computing inclusive B--meson decay spectra. This progress was achieved by DGE, which replaces the ``shape function'' by resummation of the perturbative expansion and parametrization of the dominant non-perturbative power corrections.
This framework provides the most accurate predictions for the experimentally relevant BF's in inclusive decays. It has passed two crucial tests in comparison with data:
\begin{itemize}
\item{} The consistency between the extracted values of {  $|V_{\rm ub}|$} from all available {  $\bar{B}\to X_u l \bar{\nu}$} measurements with a variety of cuts, as summarized in Fig.~\ref{fig:HFAG}.
\item{} The agreement with the measurements of the first and second moments of the photon energy in {$\bar{B}\to X_s \gamma$} for all cuts, as shown in Fig.~\ref{fig:cut_moments}.
\end{itemize}
So far the theoretical predictions for the spectra~\cite{Andersen:2005bj,Andersen:2005mj,Andersen:2006hr} have been based on resummed perturbation theory for the on-shell heavy quark decay, while non-perturbative Fermi--motion effects have only been considered as part of the error analysis.
As discussed in Sec.~\ref{sec:DGE}, a main advantage of DGE is the direct link it makes between the perturbative and non-perturbative sides of the problem: definite predictions for the parametric form of the power corrections emerge from the resummation formalism. By using experimental data to determine the power corrections, the full predictive power of DGE can be put to use.

In both $\bar{B}\to X_s \gamma$ and the  $\bar{B}\to X_u l \bar{\nu}$  the dominant theoretical uncertainties (aside from the sensitivity to the b quark mass)
are still associate with perturbative corrections. In both cases, there has been significant progress on NNLO calculations but major challenges on the way to this goal still lie ahead.

In $\bar{B}\to X_s \gamma$, despite tremendous progress in NNLO calculations made by several groups, the theoretical uncertainty in the total BF is still larger than the experimental one. A significant reduction of the uncertainty is expected upon completion of the NNLO calculation.
As in other cases, also here: the measurements are (so far) consistent with the SM predictions --- compare (\ref{radiative_exp}) with~(\ref{radiative_result}).

In {$\bar{B}\to X_u l \bar{\nu}$}, the \emph{fully differential width} has recently been computed to all orders in the large--$\beta_0$ limit~\cite{Gambino:2006wk}. By matching the DGE spectrum to the newly available {$\beta_0\alpha_s^2$} result we expect to improve the determination of $|V_{\rm ub}|$ from all kinematic cuts. Further reduction in the theoretical uncertainty can be achieved by more accurate determinations of $m_b$ and constraints on power corrections. These require dedicated comparison between theory and data for both the $\bar{B}\to X_s \gamma$ moments and for partial BF's and moments of the charmless semileptonic decay itself.

\begin{widetext}

\end{widetext}
\end{document}